\documentclass[final,5p,times,twocolumn, sort&compress]{elsarticle}

\usepackage{amssymb,color}
\usepackage{graphicx}
\usepackage{epstopdf}
\usepackage{multirow,slashbox}

\graphicspath{{figs/}}

\journal{Journal of the Mechanics and Physics of Solids}

\newcommand{\hsp}[1]{\textcolor{red}{#1}}

\begin{document}

\begin{frontmatter}



\title{A Surface Stacking Fault Energy Approach to Predicting Defect Nucleation in Surface-Dominated Nanostructures}


\author[JWJ]{Jin-Wu Jiang}

\author[AML]{Austin M. Leach}

\author[KG]{Ken Gall}

\author[HSP]{Harold S. Park\corref{cor1}}
\ead{parkhs@bu.edu}

\author[TR]{Timon Rabczuk\corref{cor2}}
\ead{timon.rabczuk@uni-weimar.de}

\cortext[cor1]{Corresponding author}
\cortext[cor2]{Corresponding author}

\address[JWJ]{Institute of Structural Mechanics, Bauhaus-University Weimar, Marienstr. 15, D-99423 Weimar, Germany}

\address[AML]{Abbott Vascular, 3200 Lakeside Drive, Santa Clara, CA 95045, USA}

\address[KG]{School of Materials Science and Engineering, Woodruff School of Mechanical Engineering, Georgia Institute of Technology, Atlanta, GA 30332, USA}

\address[HSP]{Department of Mechanical Engineering, Boston University, Boston, MA 02215, USA}

\address[TR]{Institute of Structural Mechanics, Bauhaus-University Weimar, Marienstr. 15, D-99423 Weimar, Germany}

\begin{abstract}

We present a surface stacking fault (SSF) energy approach to predicting defect nucleation from the surfaces of surface-dominated nanostructure such as FCC metal nanowires.  The approach leads to a criteria that predicts the initial yield mechanism via either slip or twinning depending on whether the unstable twinning energy or unstable slip energy is smaller as determined from the resulting SSF energy curve.  The approach is validated through a comparison between the SSF energy calculation and low-temperature classical molecular dynamics simulations of copper nanowires with different axial and transverse surface orientations, and cross sectional geometries.  We focus on the effects of the geometric cross section by studying the transition from slip to twinning previously predicted in moving from a square to rectangular cross section for $\langle100\rangle/\{100\}$ nanowires, and also for moving from a rhombic to truncated rhombic cross sectional geometry for $\langle110\rangle$ nanowires.  \hsp{We also provide the important demonstration that the criteria is able to predict the correct deformation mechanism when full dislocation slip is considered concurrently with partial dislocation slip and twinning.  This is done in the context of rhombic aluminum nanowires which do not show a tensile reorientation due to full dislocation slip.}  We show that the SSF energy criteria successfully predicts the initial mode of surface-nucleated plasticity at low temperature, while also discussing the effects of strain and temperature on the applicability of the criterion.  

\end{abstract}


\end{frontmatter}


\section{Introduction}

One of the fundamental mechanical properties of a material is its yield strength, or the maximum amount of load a material can take before sustaining irreversible plastic deformation, and eventually failure or fracture.  In metals, because the unit plastic deformation mechanism is the dislocation, researchers have focused on developing models to predict when a metal will yield via dislocation nucleation, and therefore what the yield strength of the metal will be.  

Seminal work towards this end was performed by~\citet{vitekPSS1966,vitekPM1968}, who formulated a curve of potential energy versus displacement known as the generalized stacking fault (GSF) curve.  The curve typically takes a nearly sinusoidal shape~\cite{zimmermanMSMSE2000}, in which the energy barrier that must be crossed in order for defect nucleation to occur is called the unstable stacking fault energy $\gamma_{us}$ (see Fig.~\ref{fig_gsf_bulk}). Upon crossing the $\gamma_{us}$ barrier, an equilibrium lattice spacing that does not correspond to the bulk equilibrium structure can be found, which corresponds to an intrinsic stacking fault (ISF).  The resulting excess energy as compared to the bulk equilibrium for this configuration is called the intrinsic stacking fault energy $\gamma_{isf}$.  In the case of face-centered cubic (FCC) metals, the displacement axis of the GSF curve is along the $\langle112\rangle$ direction as this is the slip direction for partial dislocations in an FCC crystal.  

The ideas underlying the GSF have been utilized for various predictions regarding nanoscale plasticity over the years, and have also formed the basis for recent theories used to predict whether defect nucleation via twinning or slip will occur.  These theories originate from the seminal work of~\citet{riceJMPS1992}, who formulated a model of dislocation nucleation from a crack tip.  Specifically, Rice considered the competition between crack tip yielding via dislocation nucleation as compared to brittle fracture by comparing the energy needed to create new fracture surfaces to the energy barrier to be overcome ($\gamma_{us}$) in order for ductile fracture to occur.   Rice's ideas were extended by~\citet{tadmorJMPS2004} and~\citet{bernsteinPRB2004} to consider the competition between dislocation nucleation and twinning at a crack tip.  However, these theories all consider bulk materials, and do not include surface effects, which are critical for plasticity in low-dimensional nanostructures such as nanowires~\cite{parkJMPS2006,parkMRS2009,weinbergerJMC2012}.  Other theories concerning the twinning versus slip competition in nanowires have recently been summarized in the excellent review of~\citet{weinbergerJMC2012}.

There has been considerable success, as demonstrated above, in applying the GSF to predicting, at an atomistic level, defect nucleation in bulk materials.  However, the validity of the GSF for predicting defect nucleation from the surfaces of nanostructures such as nanowires has not been demonstrated to-date.  

Nanowires are an important class of low-dimensional nanostructures that have been extensively studied in the past ten or so years~\cite{xiaAM2003}.  Because they are one-dimensional in nature, and because they are characterized by a large surface area to volume ratio, their transverse surfaces have been found to play a critical role in controlling their operant inelastic deformation mechanisms.
The deformation mechanisms of metal nanowires have been studied in detail recently, primarily through classical molecular dynamics (MD) simulations.  While more extensive reviews on the deformation and plasticity of metal nanowires have recently appeared~\cite{parkMRS2009,weinbergerJMC2012}, we highlight here those works that are particularly relevant to the present work, i.e. those which demonstrate the unexpected surface-mediated inelastic deformation mechanisms in nanowires that motivate the present work.

One important finding was that the transverse surfaces have a direct, first order effect on the operant inelastic deformation mechanism~\cite{parkJMPS2006} of metal nanowires.  This work~\cite{parkJMPS2006} also demonstrated that traditional notions used to predict deformation bulk mechanisms such as the Schmid factor may not be able to capture the surface-dominated mechanisms observed in nanowires.  Another important finding was that the geometry of the nanowire cross section also has a direct effect on the operant inelastic deformation mechanism of nanowires~\cite{jiAPL2006,jiNANO2007a,leachAFM2007}.  ~\citet{jiAPL2006,jiNANO2007a} elucidated this by considering the deformation of $\langle100\rangle$ copper nanowires with cross sectional geometries varying from square to rectangular.  They found that while square nanowires yielded by nucleation and propagation of full and partial dislocation slip, rectangular nanowires yielded by nucleation and propagation of twins.  A similar conclusion was found by~\citet{leachAFM2007}, who studied the effects of truncating a rhombic cross section on the deformation mechanisms of silver nanowires.  While fully rhombic cross section $\langle110\rangle$ nanowires were observed to reorient in tension via nucleation and propagation of twins to a $\langle100\rangle$ orientation~\cite{parkPRL2005,liangNL2005}, truncated rhombic nanowires were found to fail via nucleation and propagation of partial dislocations, while pentagon cross section nanowires were also found to fail via nucleation and propagation of partial dislocations.

These MD studies collectively demonstrate that a predictive model of surface-nucleated incipient plasticity should account for loading mode (i.e. tension or compression), transverse surface orientation, nanowire axial orientation and the geometry of the nanowire cross section.  It is worth noting that while the experimental literature on deformation mechanisms in metal nanowires is quite sparse compared to the fully populated literature on MD simulations, evidence of the effects discussed above has emerged.  For example, the brittle nature of deformation and fracture of pentagonal silver nanowires was recently demonstrated by~\citet{zhuPRB2012}.  Similarly, rhombic cross section $\langle110\rangle$ gold nanowires were observed to undergo long-ranged twin migration and propagation, enabling a tensile-stress-induced reorientation to square cross section $\langle100\rangle$ nanowires~\cite{seoNL2011}.  In contrast, pristine hexagonal cross section copper nanowires were observed to undergo brittle fracture by~\citet{richterNL2009}, which suggests that the cross sectional geometry of the nanowires is in fact a critical factor controlling the inelastic deformation mechanism.  

Outside of classical MD simulations, there have also been atomistic approaches based on classical stability theory, in which defect nucleation at the surfaces of nanostructures occurs when either the determinant of the Hessian becomes negative~\cite{dmitrievAM2005}, or when the change in potential energy from one configuration to another decreases~\cite{yunIJSS2011}.  However, both of these approaches were for defect nucleation from infinitely large surfaces attached to semi-infinite bodies, and therefore may not hold predictive power for defect nucleation from the surfaces of one-dimensional nanowires.  Other researchers have also utilized energy landscape exploration techniques to investigate the nucleation barriers that must be overcome for specific plastic deformation mechanisms to occur~\cite{zhuPRL2008,weinbergerJMPS2012,ryuPNAS2011}.  These approaches, while insightful, have also not yielded a generic and predictive capability that correlates the observed incipient plastic deformation mechanism to the nanowire size, cross sectional geometry, axial and surface orientation.

Therefore, we present in this work a generalized stacking fault criteria for surfaces.  The work is based upon the classical GSF approach of~\citet{vitekPSS1966,vitekPM1968}, though applied to finite domain nanostructures such as nanowires; we note that the the bulk GSF-like criteria of~\citet{bernsteinPRB2004} was used by~\citet{liangPRB2006} to explain the deformation mechanisms of rhombic $\langle110\rangle/\{111\}$ NWs, though again, the criteria of~\citet{bernsteinPRB2004} did not include surface effects.   Ideas very similar to the ones presented in this manuscript were discussed by~\citet{leachTHESIS} in a Ph.D. thesis, though no work along these lines was published. We demonstrate that because this approach accounts for all features discussed above, i.e. loading mode, transverse surface orientation, axial orientation and cross sectional geometry, it is applicable to a wide range of metal nanowire geometries and loading modes.  We illustrate the proposed approach as compared to companion benchmark MD simulations.  

\section{Methodology}

For all MD simulations performed in this work, the interaction between copper atoms is described using the embedded atom method (EAM)~\cite{foilesPRB1986}, using the potential developed by Mishin {\it et al.}~\cite{mishinPRB2001}. In the EAM, the total energy of the system is expressed in terms of the summation over each atom $i$,
\begin{eqnarray}
E=\sum_{i=1}^{N}F_{i}(\rho_{i}) + \sum_{i< j}^{N}\phi_{ij}(r_{ij}),
\label{eq_eam}
\end{eqnarray}
where $N$ is the total number of atoms.  Atom $i$ is embedded in its position by the first `embedding' term with respect to the electron density $\rho_{i}$ at this position. The second term provides the repulsive force between atom $i$ and its neighbors, which helps to stabilize the whole structure.  The EAM potential of Mishin {\it et al.}~\cite{mishinPRB2001} was chosen due to its accuracy in capturing the stacking fault and surface energies of copper.  

All MD simulations were performed using the publicly available simulation code LAMMPS~\cite{plimptonJCP1995,LAMMPS}, while the OVITO package was used for visualization~\cite{stukowskiA2010}.  In the MD simulation, the system is first thermalized at a constant temperature by the N\'ose-Hoover~\cite{noseJCP1984,hooverPRA1985} heat bath for 100~{ps}. Free boundary conditions were applied along the two lateral directions of the nanowire by introducing a large vacancy space in the simulation box along these two directions.  Along the axial direction, one end of the nanowire was fixed while the other was free.  Tensile strain was then applied, similar to previous works~\cite{parkPRB2005}, by adding a ramp velocity that scaled linearly from zero at the fixed end to a maximum value at the free end. The added velocity profile results in a strain rate of $\dot{\epsilon}=10^{9}$~s$^{-1}$, which is a typical value in MD simulations, and which does not preclude the generation of twins, full or partial dislocations, as shown in previous works~\cite{parkJMPS2006,jiAPL2006,jiNANO2007a,leachAFM2007,parkPRL2005,liangNL2005}.

\subsection{Bulk GSF Energy Curve Calculation}

\begin{figure}\begin{center}
\includegraphics[scale=0.5]{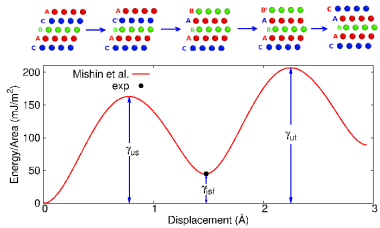}
\caption{(Color online) The GSF energy curve for bulk FCC copper for slip on a $\{111\}$ plane along the $\langle112\rangle$ direction. The experimental value for the intrinsic stacking fault energy is shown as a dot. The five extreme points are according to the five corresponding configurations shown in the top inset. The unstable stacking fault energy ($\gamma_{\rm us}$), the intrinsic stacking fault energy ($\gamma_{\rm isf}$) and the unstable twinning fault energy ($\gamma_{\rm ut}$) are displayed.} 
\label{fig_gsf_bulk} \end{center} \end{figure}

For bulk FCC copper, the GSF energy curve is calculated using standard procedures~\cite{zimmermanMSMSE2000}.  Specifically, the simulation block is rectangular with its three orthogonal coordinate directions along the $\langle111\rangle$, $\langle110\rangle$, and $\langle112\rangle$ lattice directions. Periodic boundary conditions are applied in the $\langle110\rangle$ and $\langle112\rangle$ directions, while free boundary conditions are applied in the $\langle111\rangle$ direction. The structure is divided into two halves by a $\{111\}$ plane. These two halves are sheared with respect to each other along the $\langle112\rangle$ direction using small displacement increments.  The total potential energy ($E_{\rm tot}$) of the structure is recorded at each shearing displacement, while the atoms are allowed to relax in the $\langle111\rangle$ direction during the shearing process. The GSF energy is then calculated from $(E_{\rm tot}-E_{\rm tot}^{0})/A_{\{111\}}$, where $(E_{\rm tot}-E_{\rm tot}^{0})$ is the change in the total potential energy due to the relative displacement between two halves and $A_{\{111\}}$ is the area of the shearing $\{111\}$ plane. 

Fig.~\ref{fig_gsf_bulk} shows the GSF energy curve for bulk copper. The top inset in the figure demonstrates the evolution of the atomic configuration at the shearing interface during the relative displacement of the two halves of the crystal. An intrinsic stacking fault (ISF) occurs at the partial Burgers vector $\vec{b}_{p}$ with $b_{p}=a/\sqrt{6}$, while an unstable stacking fault (US) occurs at the displacement of $\vec{b}_{p}/2$. Subsequent shearing results in the second peak in the bulk GSF energy curve, which corresponds to the unstable twinning (UT) energy.  As expected, good agreement in the ISF energy $\gamma_{isf}$ is obtained between the Mishin {\it et al.}~\cite{mishinPRB2001} potential (44.5 mJ/m$^2$) and experiments (45.0 mJ/m$^{2}$)~\cite{hirth}.

\subsection{Surface Stacking Fault Energy Curve Calculation}

\begin{figure}\begin{center}
\includegraphics[scale=0.38]{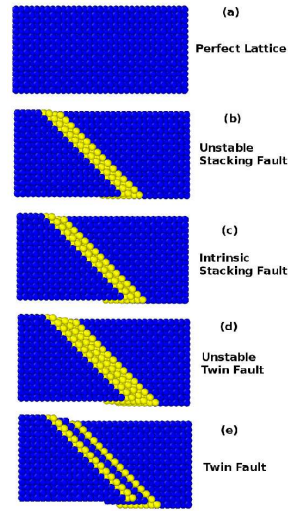}
\caption{(Color online) The formation of a twinning fault in a $\langle100\rangle/\{100\}$ CuNW, where the $\{111\}$ plane slips in the $\langle112\rangle$ direction.}
\label{fig_gsf_nw_twin} \end{center} \end{figure}

The surface stacking fault (SSF) energy curve for unstrained nanowires can be calculated via the following two steps. First, the nanowire is relaxed to a minimum energy configuration.  During this process, surface atoms contract inwards towards the bulk atoms to maximize their electron density.  Second, the SSF energy curve for the twinning fault of the nanowire can then be obtained through the process shown in Fig.~\ref{fig_gsf_nw_twin}, where the $\langle100\rangle/\{100\}$ CuNW in (a) is sheared rigidly along a $\{111\}$ plane in the $\langle112\rangle$ direction.  Unlike in the bulk case, the two blocks of atoms comprising the nanowire are not allowed to relax in the nanowire case.  This is primarily due to the difficulty in constraining the nanowire relaxation to occur only in the $\langle111\rangle$ direction, as in the bulk case.  However, as we demonstrate below, this small approximation does not degrade the accuracy or utility of the SSF curves.  As the strain increases, the configuration of the unstable stacking fault is reached in (b), followed by the formation of an ISF in (c).  After the first ISF is formed, the deformation continues to shear a neighboring $\{111\}$ plane in the same $\langle112\rangle$ direction, resulting in the formation of the second unstable twinning fault seen in (d). Finally, the formation process finishes with the twinning fault in (e).  Similar to the bulk case, the SSF energy for nanowire here is then calculated from $(E_{\rm tot}-E_{\rm tot}^{0})/A_{\{111\}}$.

We also calculate the SSF energy curves for axially strained nanowires, for reasons that we explain in detail in the discussion of the numerical examples.  For these SSF curves, the nanowire is strained axially in increments of 0.1\%, where the atoms are allowed to relax after the two ends of the nanowire are fixed at the new length.  From these energy minimizing positions, the shearing process described above and illustrated in Fig. \ref{fig_gsf_nw_twin} is performed to extract the SSF energy curve for the prescribed axial strain.

\begin{figure}\begin{center}
\includegraphics[scale=0.38]{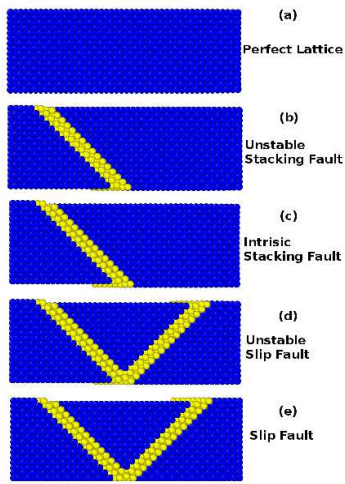}
\caption{(Color online) The formation of a slip fault in a $\langle100\rangle/\{100\}$ CuNW.}
\label{fig_gsf_nw_slip} \end{center} \end{figure}

Fig.~\ref{fig_gsf_nw_slip} demonstrates the formation process of a slip fault. The production of the first ISF from (a) to (c) is the same as the formation process of the twinning fault shown in Fig.~\ref{fig_gsf_nw_twin}, so the SSF energy curve are the same for twinning and slip faults up to the occurrence of the first ISF in Fig.~\ref{fig_gsf_nw_slip}(c). The slip formation continues by generating a second ISF, where the nanowire is sheared along another $\{111\}$ plane along its corresponding $\langle112\rangle$ direction. The slip fault is formed in (e). The SSF energy for the slip fault is also calculated from the change of the total potential energy in the structure.

\begin{figure}\begin{center}
\includegraphics[scale=0.7]{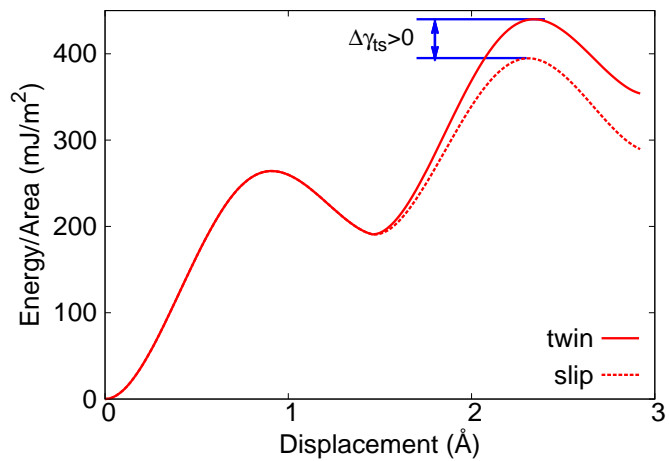}
\caption{(Color online) The SSF energy curve for a CuNW of dimension $40\times4\times4$ lattice constants. The energy barrier ($\Delta \gamma_{ts}$) shown in the figure is the energy difference between the second peaks of the unstable twinning and slip SSF curves.  Therefore, for this specific example, our criteria predicts that slip deformation should be the preferred deformation pathway in this CuNW, because the energy barrier ($\Delta\gamma_{ts}$) is positive.}
\label{fig_gsf_cubic_40_4_4} \end{center} \end{figure}

The SSF energy curves for twinning and slip deformation pathways are the same up to the emergence of the first ISF. Afterwards, these two SSF energy curves start to differ. At the second unstable stacking fault, if the twinning SSF energy is lower than that of the slip deformation, then the twinning deformation will be the energetically favored pathway during the inelastic deformation of this system. Otherwise, slip will be the energetically favored pathway.  

Thus, the basis of the SSF-based defect prediction criteria is to compare the energetic cost to nucleating a slip defect, or a twinning defect as calculated using the SSF energy curve.  If the energetic cost for slip is larger than twinning, then the prediction is that twinning will occur first.  Conversely, if the energetic cost for twinning is larger than slip, then the prediction is that slip will occur first.  In this way, the energy barrier between the twinning and the slip deformation in the SSF energy curve can be used as a criteria to predict the energetically favorable plasticity mechanism during the deformation of the nanowire.

The criteria is explicitly demonstrated in Fig.~\ref{fig_gsf_cubic_40_4_4} for a CuNW of dimension $40\times4\times4$ lattice constants, where the SSF energy curves for twinning and slip are compared.  In Fig.~\ref{fig_gsf_cubic_40_4_4}, the energy barrier ($\Delta\gamma_{ts}$), which is the energy difference between the second peaks of the unstable twinning and slip SSF curves, is also displayed. Therefore, for this specific example, the criteria predicts that the initial plastic deformation event will be slip deformation, as the energy barrier $\Delta\gamma_{ts}$ has a positive value.  

We should note that the energetics of the SSF deformation path depends on where along the nanowire the second partial occurs. In some cases, the energy will be lower if the second partial occurs close to the first partial dislocation. Otherwise, if the two partials in the slip process occur at different places along the nanowire, then the SSF energy curve for slip deformation will be the same as for the twinning deformation.

\section{Results}

As shown below, we consider two numerical examples.  The first considers a tensile loading-induced transition from slip to twinning that was previously predicted in CuNWs~\cite{jiAPL2006} when the cross sectional geometry changed from square to rectangular.  The second considers a tensile loading-induced transition from twinning to slip that was previously predicted in AgNWs~\cite{leachAFM2007} in going from a rhombic to truncated rhombic cross sectional geometry.

We emphasize that these examples were not chosen arbitrarily.  Specifically, it has been established in the literature that defect nucleation in nanowires is dependent, to varying degrees, on the Schmid factor, the (bulk) GSF energy, and the orientation of the transverse surfaces.  In fact, a very lucid and detailed analysis of MD simulations of nanowire plasticity, and the ability of these three key factors in predicting the observed deformation mechanisms, was recently performed by~\citet{weinbergerJMC2012}. In that work, it was noted that while it is known that the three factors are important, a predictive model that incorporated all three effects is currently lacking.

In view of this, the examples discussed below were chosen specifically because they show a transition in the operant deformation mechanism that cannot be captured by bulk deformation measures such as the Schmid factor, which is purely geometric and does not capture surface effects~\cite{parkJMPS2006}, and the bulk GSF energy, which again neglects surface effects.  The effects of side surface are important, but only seen in the case of the rhombic NWs, which exhibit a change in the initial surface structure with increasing truncation of the cross sectional geometry.  As will be illustrated below, the SSF model does account for changes in initial surface structure, but because it combines a variety of effects (changes in cross sectional geometry, surface orientation, axial orientation and size) into a single constant (the SSF curve), further investigation will be required to delineate each of these individual effects on the observed transition in deformation mechanism.

\subsection{Square to Rectangular NWs}

The rationale for the simulations in this section are to investigate the ability of the proposed SSF criteria to capture the transition from slip to twinning-dominated deformation that was observed in earlier MD simulations~\cite{jiAPL2006} when the cross section of $\langle100\rangle/\{100\}$ copper nanowires (CuNWs) changed from square to rectangular.  We perform such simulations, while calculating the SSF energy curves for both twinning and slip deformation pathways. These SSF energy curves are then used to predict for the favored deformation pathway in the CuNWs.  Concurrent low-temperature (0.1K) MD simulations are performed to monitor the deformation pathway during the inelastic deformation of the CuNWs under tension.  We will validate the SSF criteria by comparing the deformation pathway predicted by the SSF energy calculations to that observed in the MD simulations.

\begin{figure}\begin{center}
\includegraphics[scale=0.65]{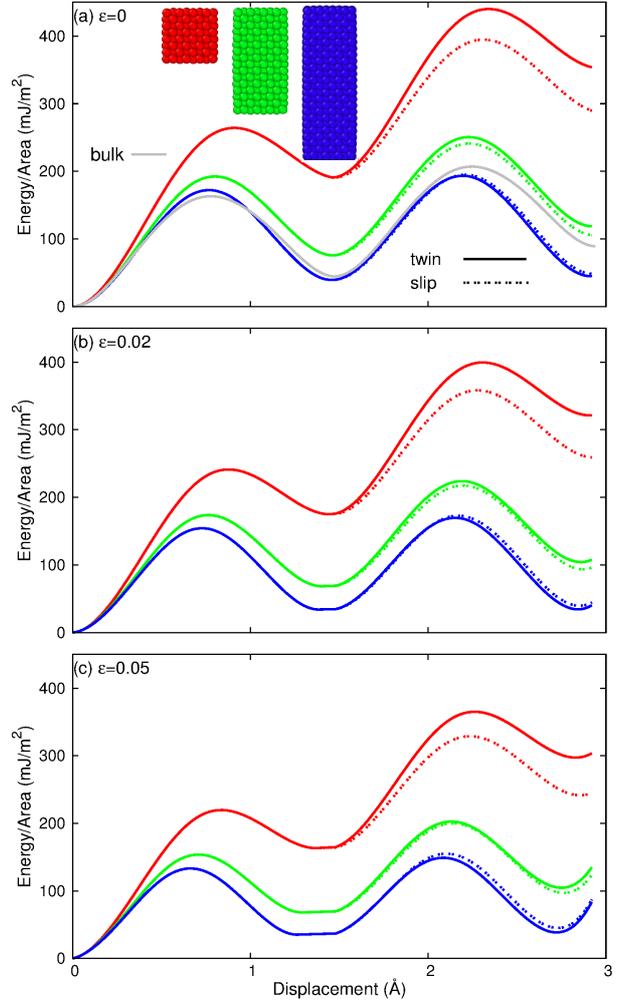}
\caption{(Color online) The SSF energy curve for three CuNWs with size $(l_{x}, l_{y}, l_{z})$=(40, 4, 4), (40, 4, 8), and (40, 4, 12). Curves for twinning and slip deformation are displayed by solid and dashed lines, respectively. Lines are colored according to the color of the cross section in the inset in (a). The SSF energy for structures with strains 0.0, 0.02, and 0.05 are shown in panels (a), (b), and (c). The bulk GSF energy curve is also shown in (a) for comparison.}
\label{fig_gsf_square_to_rectangular} \end{center} \end{figure}

The three nanowires studied had size $(l_{x}, l_{y}, l_{z})$=(40, 4, 4), (40, 4, 8), and (40, 4, 12) cubic lattice units (CLU), where the CLU=$a=3.615$~{\AA} is the lattice constant of FCC copper. All three CuNWs had the same axial orientation along the $\langle100\rangle$ direction, with four $\{100\}$ side surfaces. Fig.~\ref{fig_gsf_square_to_rectangular}~(a) shows the SSF energy curves of the undeformed CuNWs.  For the two CuNWs with $l_{z}=$ 4 and 8, the deformation energy required for slip is lower than that for the twinning. For $l_{z}=12$, the SSF energy curves for the twinning and slip pathways are almost indistinguishable, which implies that either twinning or slip could occur.

\begin{figure}\begin{center}
\includegraphics[scale=0.7]{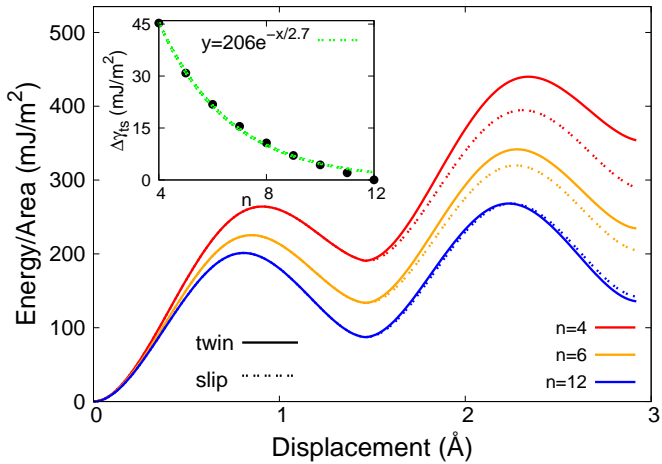}
\caption{(Color online) Size dependence for the square CuNWs $(40, n, n)$ with $n=$ 4, ..., 12. Inset shows the energy barrier between twinning and slip deformation pathways; i.e the difference between the unstable twinning energy and the unstable slip energy in the second peak of the SSF energy curve.}
\label{fig_gsf_square_size} \end{center} \end{figure}

We also examined size effects on the SSF curve in square CuNWs of dimension $(40, n, n)$ with $n=$ 4, ..., 12, with the results summarized in Fig.~\ref{fig_gsf_square_size}. Our calculations show that for all of this set of square cross-section CuNWs, the deformation mechanism is always predicted to be slip, so the preferred deformation pathway is size-independent. However, with increasing $n$, the energy barrier difference between slip and twinning deformation decays exponentially, which suggests that an exponential decay is a characteristic property for surface or edge effects with regards to the deformation mechanism.

We also show the SSF energy curves for all the three CuNWs with strain $\epsilon=0.02$ and 0.05 in Figs.~\ref{fig_gsf_square_to_rectangular}~(b) and (c), where we emphasize that these strains are below the yield strain of all CuNWs. By performing the SSF calculation on the tensile-strained nanowires, it is clear that the strain has no effect on the square CuNW (40,4,4), where slip remains the preferred deformation mechanism.  

In contrast, the difference in the energy barrier for twinning and slip in the CuNW (40, 4, 8) becomes very small at $\epsilon=0.05$, with the SSF energy for slip being slightly lower than twinning.  
We have also calculated the SSF energy curve for CuNW (40, 4, 8) with even larger tensile strains, and observed similar trends, i.e that the energy for slip is always very close to but slightly lower that for twinning.  However, as the cross sectional geometry becomes even more rectangular as for the CuNW (40,4,12) case, it is seen that the twinning pathway energy barrier becomes progressively smaller as strain increases than that for slip.  These results are in agreement with the MD simulation results of~\citet{jiAPL2006}, who predicted the slip to twinning transition as the nanowire cross sectional geometry transitioned from square to rectangular.  The energy barriers between twinning and slip deformation for the square to rectangular CuNWs are summarized in Table~\ref{tab_energy_barrier}.

\begin{table*}
\caption{Structure and strain ($\epsilon$) dependence for the energy barrier ($\Delta \gamma_{ts}$) between twinning and slip deformation pathway, i.e the energy difference between the second peak in the SSF energy curve for the twinning and slip deformation. Positive $\Delta E$ predicts slip deformation pathway; while negative energy barrier corresponds to twinning deformation pathway.}
\label{tab_energy_barrier}
\begin{tabular*}{\textwidth}{@{\extracolsep{\fill}}|c|c|c|c|c|c|c|c|c|}
\hline
structure & \multicolumn{3}{c|}{square to rectangular $(40, 4, n)$} & \multicolumn{5}{c|}{rhombic with truncation level $n$} \\
\hline
{\backslashbox{$\epsilon$}{n}} & 4 & 8 & 12 &  0 & 1 & 2 & 3 & 4 \\
\hline
0.0 & 45.3 & 9.6 & -1.5 & -17.2 & -9.3 & 0.1 & 24.6 & 26.1 \\
\hline
0.02 & 41.2 & 6.6 & -3.2 & -19.5 & -12.1 & -4.2 & 18.7 & 21.8 \\
\hline
0.05 & 36.2 & 2.6 & -6.0 & -23.3 & -17.2 & -10.1 & -7.7 & 6.4 \\
\hline
\end{tabular*}
\end{table*}

\begin{figure}\begin{center}
\includegraphics[scale=0.55]{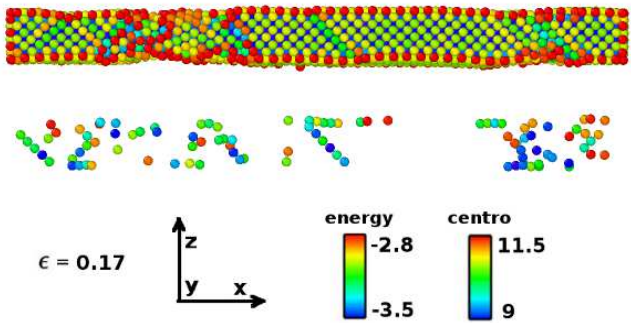}
\caption{(Color online) The configuration for the CuNW (40, 4, 4) from MD simulation at 0.1~{K} with strain $\epsilon=0.17$. Atoms are colored by potential energy in the top image and by the centrosymmetry parameter in the bottom image.}
\label{fig_md_cubic_40_4_4_0.1K} \end{center} \end{figure}

We verify the predictions of the SSF curves in Fig.~\ref{fig_gsf_square_to_rectangular} by performing MD simulations of the tensile deformation of CuNWs.  Because of the strain-induced evolution in the predicted deformation mechanism seen in Fig.~\ref{fig_gsf_square_to_rectangular}, we use the SSF energy calculated from a strained CuNW in Fig.~\ref{fig_gsf_square_to_rectangular}(c) to predict the deformation pathway we expect to see in the MD simulations.  From Fig.~\ref{fig_gsf_square_to_rectangular}~(c), the energetically preferred deformations predicted by the SSF energy are slip, slip/twinning, and twinning for the three CuNWs with $l_{z}=$ 4, 8, and 12. For $l_{z}=8$, two deformation pathways can occur simultaneously due to their close SSF energy curves, and thus our SSF calculation predicts a transition from slip to twinning deformation in CuNWs $(l_{x}, l_{y}, l_{z})$ with increasing width to height ratio $l_{z}/l_{y}$. 

Fig.~\ref{fig_md_cubic_40_4_4_0.1K} shows the configuration of the CuNW (40, 4, 4) under strain $\epsilon=0.17$ at 0.1~{K}. The top image is colored according to the potential energy per atom, while the bottom image is colored by the centrosymmetry parameter. Only atoms with centrosymmetry parameters within the color bar are shown in the bottom image. No twinning deformation is observed in the deformed structure, while slip deformations can be observed.  This MD simulation result is therefore consistent with the prediction by the SSF energy in Fig.~\ref{fig_gsf_square_to_rectangular}~(c).

\begin{figure}\begin{center}
\includegraphics[scale=0.5]{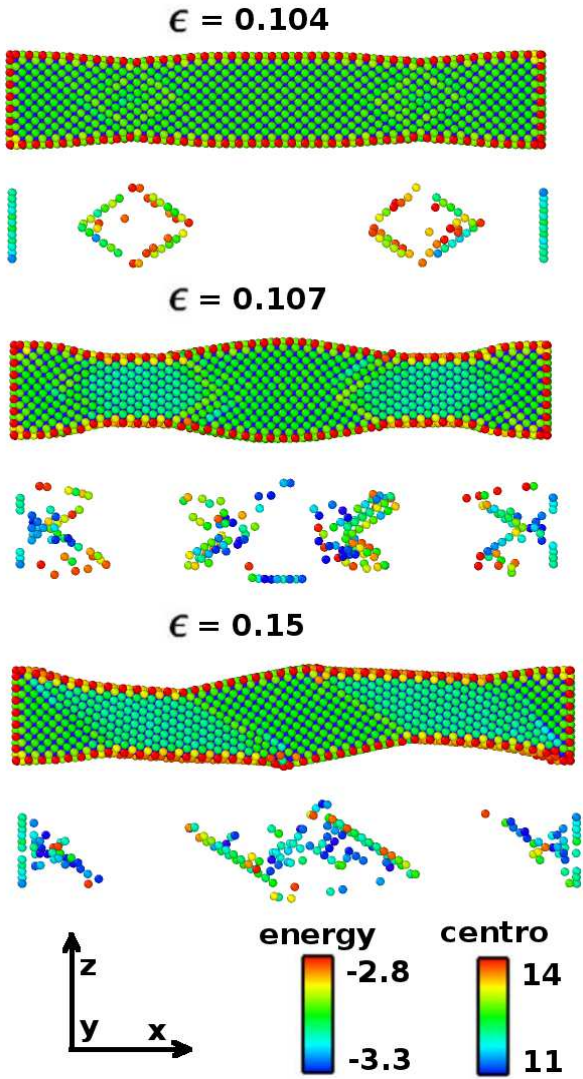}
\caption{(Color online) The evolution of the configuration for CuNW (40, 4, 8) at 0.1~{K} with increasing tensile strain. The top image in each figure is colored according to potential energy while the bottom image is colored by the centrosymmetry parameter. At $\epsilon=0.104$, partial dislocations from the top and bottom surfaces form two rhombic shapes. These rhombic shapes propagate along the CuNW under increasing tensile strain before $\epsilon<0.107$ and turn into twinning deformation at $\epsilon=0.15$.}
\label{fig_md_cubic_40_4_8_0.1K} \end{center} \end{figure}

Fig.~\ref{fig_md_cubic_40_4_8_0.1K} shows the evolution of the configuration for the CuNW (40, 4, 8) from MD simulation at 0.1~{K}. With increasing tension, the CuNW starts to yield at $\epsilon=0.104$, where the initial yield mechanism is partial dislocation slip from the top and bottom surfaces that meet at the center of the nanowire to form a rhombic dislocation structure as illustrated by the bottom image for $\epsilon=0.104$. Under slightly higher tension, $\epsilon=0.107$, the rhombic deformation shapes break open and propagate along the axial direction of the nanowire. At $\epsilon=0.15$, a transition from slip to twinning is observed.  These MD simulation results are consistent with the SSF energy prediction in Fig.~\ref{fig_gsf_square_to_rectangular}~(c).

\begin{figure}\begin{center}
\includegraphics[scale=0.45]{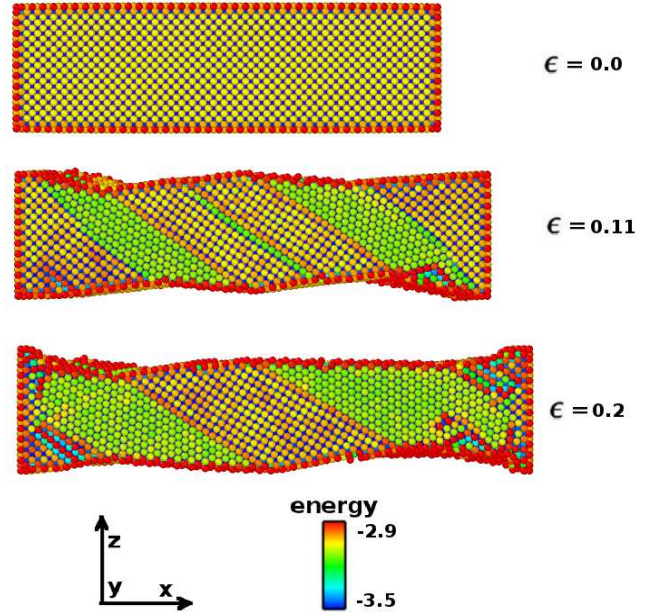}
\caption{(Color online) Tensile deformation of the CuNW (40, 4, 12) at 0.1~{K}. Twinning is the only observed plastic deformation mechanism.}
\label{fig_md_cubic_40_4_12_0.1K} \end{center} \end{figure}

Fig.~\ref{fig_md_cubic_40_4_12_0.1K} shows the deformation of the CuNW (40, 4, 12) under tensile loading at 0.1~{K}. Twinning is found to be the sole deformation mechanism throughout the MD simulation. Specifically, there are several observable twin boundaries in separate regions of the nanowire at $\epsilon=0.11$. With increasing strain $\epsilon=0.2$, these twin boundaries separate to form larger twinned regions in the nanowire.  The MD simulations again verify the prediction from the SSF energy in Fig.~\ref{fig_gsf_square_to_rectangular}~(c).

To explain the observed twinning, we note that all transverse surfaces are initially $\{100\}$. If the cross section is square, there is no asymmetric (surface) energetic driving force; instead, slip systems on alternating $\{111\}$ planes are activated which keeps the cross section square.  In contrast, wires with a rectangular cross section have two transverse $\{100\}$ surfaces that have a majority of the available transverse surface area.  We note that the twinning occurs on the larger $\{100\}$ surfaces, which allows both to reduce their surface area by reorienting to the close-packed $\{111\}$ surface. The asymmetry of the cross sectional geometry thus results in a sufficient surface energetic driving force for the larger surfaces to reduce their area and therefore their energy through reorientation via twinning-dominated deformation.

Overall, we have shown in this section that the SSF criteria can correctly predict the previously observed transition from slip to twinning in CuNWs as the cross sectional geometry evolves from square to rectangular, particularly if strain effects on the SSF curve are accounted for.

\subsection{Rhombic NWs}

\begin{figure}\begin{center}
\includegraphics[scale=0.33]{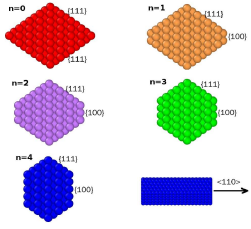}
\caption{(Color online) Cross section of the rhombic CuNW with truncation level $n=$ 0, 1, 2, 3, and 4. The side view in the last image displays the $\langle110\rangle$ growth direction of the CuNW. For perfect rhombic CuNW, all side surfaces are $\{111\}$ planes. Two $\{100\}$ lateral side surfaces will be exposed by truncation.}
\label{fig_cfg_rhombic} \end{center} \end{figure}

The rationale for the simulations in this section are to investigate the ability of the proposed SSF criteria to capture the transition from slip to twinning-dominated deformation that was observed in earlier MD simulations~\cite{leachAFM2007} when the cross section of $\langle110\rangle/\{111\}$ silver nanowires changed from rhombic to truncated rhombic.  We investigate this phenomenon in the present work for copper nanowires (CuNWs).  The truncations considered are shown in Fig.~\ref{fig_cfg_rhombic}, where the cross section is truncated with truncation level $n=$ 0, 1, 2, 3, and 4.  For $n=0$, all side surfaces are $\{111\}$ planes, with the length of this set of CuNWs being 10 nm.  The CuNW is truncated at the two lower apex, resulting in the exposure of two $\{100\}$ side surfaces. With increasing truncation level $n$, the area of the exposed $\{100\}$ surfaces increases.  In the following, we will first present the SSF energy calculation for this set of CuNWs, followed by the MD simulation results.

\begin{figure}\begin{center}
\includegraphics[scale=0.8]{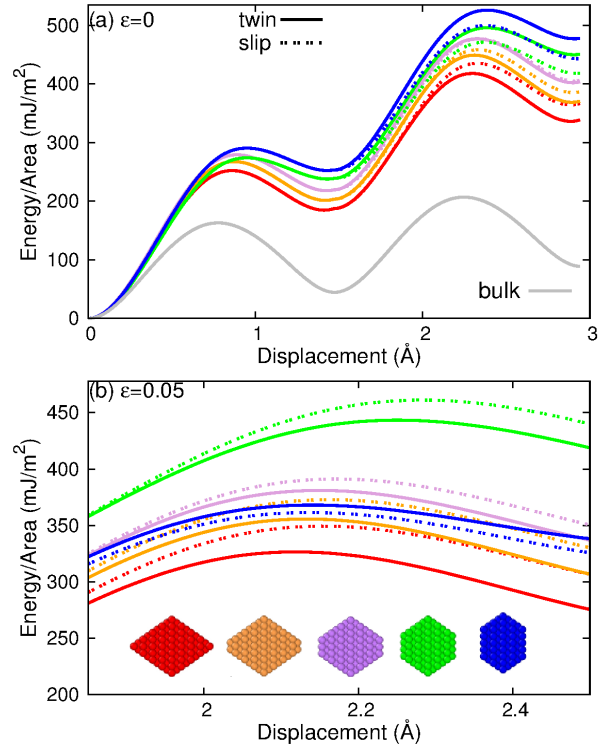}
\caption{(Color online) The SSF energy curves for truncated rhombic CuNWs. Curves are displayed by solid lines for twinning and by dashed lines for slip deformation. Lines are colored according to the color of the cross section in the inset in (b). Panel (b) shows the SSF energy curve around the second maximum region for nanowires with strain $\epsilon=0.05$. The GSF energy curve for bulk system is also shown in (a) for comparison.}
\label{fig_gsf_rhombic} \end{center} \end{figure}

The SSF energy curves for all five truncated rhombic CuNWs are shown in Fig.~\ref{fig_gsf_rhombic}~(a) for structures without strain. The curves for twinning are displayed by solid lines while the curves for slip are displayed by dashed lines. All curves are colored according to the corresponding cross section as shown in the inset of Fig.~\ref{fig_gsf_rhombic}~(a). With increasing truncation levels $n$ from 0 to 4, there is a transition in the relationship between the deformation energy of twinning and slip pathways. The transition happens at $n=2$, where the energy barrier difference between twinning and slip pathways is almost zero. 

As discussed above for the square to rectangular cross section nanowires, accounting for the effects of strain is important to accurately apply the SSF energy criteria. We thus show in Fig.~\ref{fig_gsf_rhombic}~(b) the SSF energy for these five CuNWs under tensile strain $\epsilon=0.05$. For clarity, only curves around the second extreme values in Fig.~\ref{fig_gsf_rhombic}(a) are shown. We will use Fig.~\ref{fig_gsf_rhombic}~(b) to predict the preferred deformation pathway for the truncated rhombic CuNWs. From these SSF energy curves, the deformation energy for twinning is lower than that of slip in rhombic CuNWs with truncated level $0\le n\le 2$. As a result, the twinning deformation is predicted to be the preferred pathway of these three CuNWs. The deformation energy for twinning is higher than that of the slip in rhombic CuNWs with truncated level $n\ge 3$. It is thus predicted that the slip deformation is the preferred pathway in these two ($n=3,4$) CuNWs.  The energy barriers between the twinning and slip deformation mechanisms for rhombic NWs with varying truncation level $n$ are summarized in Table~\ref{tab_energy_barrier}. We will perform MD simulations to show the transition from twinning to slip in the deformation pathway for this set of CuNWs with increasing $n$ from 0 to 4. Similar MD simulations can be found in the work by Leach {\it et.al}~\cite{leachAFM2007}.

\begin{figure}\begin{center}
\includegraphics[scale=0.45]{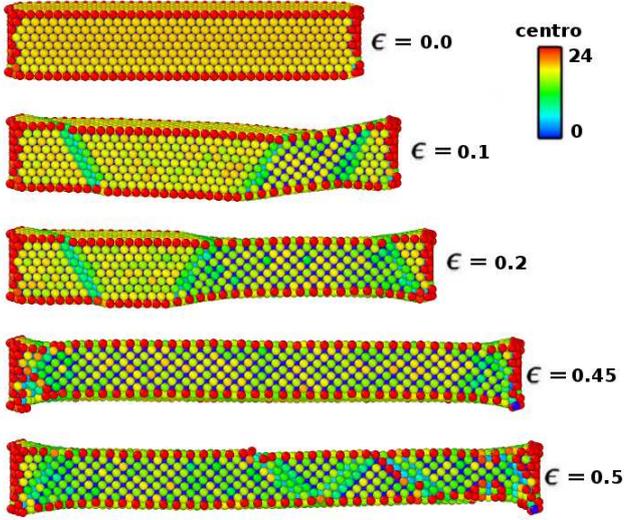}
\caption{(Color online) The evolution of the rhombic $\langle110\rangle/\{111\}$ CuNW with truncation level $n=0$ from MD simulation at 0.1~{K}. Atoms are colored by the centrosymmetry parameter.  Reorientation of the initially rhombic $\langle110\rangle$ CuNW to a square cross section $\langle100\rangle/\{100\}$ CuNW via long-range twin boundary migration is observed.}
\label{fig_md_rhombic_0_0.1K} \end{center} \end{figure}

Fig.~\ref{fig_md_rhombic_0_0.1K} shows the tensile deformation at 0.1~{K} in the MD simulation for the rhombic CuNW with truncation level $n=0$. The only observed deformation mechanism is twinning, which is generated at lower strain and then propagates along the nanowire length under further tensile loading.  The long-range propagation of the twin boundaries reorients the side surfaces from $\{111\}$ to $\{100\}$, while the axial orientation reorients from $\langle110\rangle$ to $\langle100\rangle$ with a corresponding change in cross sectional geometry from rhombic to square.  These simulation results are consistent with earlier MD simulations focused on nanowire shape memory and pseudoelasticity~\cite{parkPRL2005,parkAM2006,liangNL2005}, and more recently the work of~\citet{seoNL2011}, who gave the first experimental demonstration of such long-ranged twin boundary propagation and the subsequent reorientation.  

After the propagation of the twinning boundaries along the nanowire length, the reorientated $\langle100\rangle/\{100\}$ nanowire undergoes yielding via slip at $\epsilon=0.5$. We similarly find through the MD simulations that twinning is the only deformation mechanism observed during the tensile deformation of the rhombic CuNWs with truncated levels $n=$ 1 and 2. These MD simulation results are in good agreement with the SSF energy shown in Fig.~\ref{fig_gsf_rhombic}~(b).

\begin{figure}\begin{center}
\includegraphics[scale=0.45]{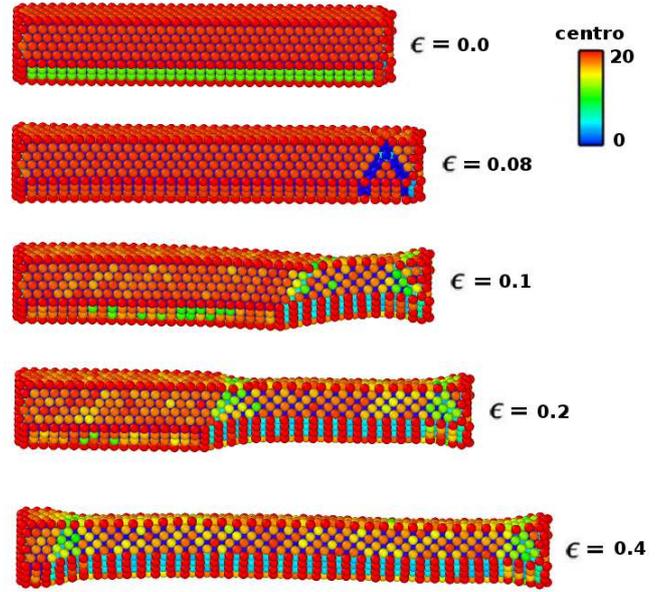}
\caption{(Color online) The evolution of the rhombic CuNW with truncation level $n=3$ from MD simulation at 0.1~{K}. Atoms are colored according to the centrosymmetry parameter. At the initial deformation stage, a slip deformation is generated at $\epsilon=0.08$, where the dislocated atoms are removed only in this particular image to clearly show the slip deformation. Under further tension, the deformation pathway switches to twinning. The twinning deformation then propagates along the entire length CuNW.}
\label{fig_md_rhombic_3_0.1K} \end{center} \end{figure}

Fig.~\ref{fig_md_rhombic_3_0.1K} shows the MD simulation for the tensile deformation of the rhombic CuNW with truncation level $n=3$ at 0.1~{K}. In the second image, copper atoms within the deformed region are removed, so that a slip deformation can be clearly observed near the right boundary at strain $\epsilon=0.08$.  With increasing tensile strain, this slip deformation turns into a twinning deformation as shown in the third image at $\epsilon=0.1$. The twinning deformation is then propagated along the nanowire and the CuNW is reorientated from a rhombic $\langle110\rangle$ CuNW into a square $\langle100\rangle$ nanowire with two \{100\} surfaces and two \{110\} surfaces.  We note that the reorientation of the side surfaces in going from the rhombic $\langle110\rangle$ to square $\langle100\rangle$ nanowire is as follows: $\{111\}\longrightarrow \{100\}$ and $\{100\}\longrightarrow \{110\}$.  

An important point to make here is that the MD simulations in Fig.~\ref{fig_md_rhombic_3_0.1K} demonstrate that while the SSF energy calculation accurately predicts the initial deformation mechanism in the nanowires, it does not yield accurate predictions of subsequent plastic events at larger strains.  This illustrates one limitation with the SSF approach, i.e. if the initial deformation mechanism is not predictive, which then results in incorrect predictions of the mechanical response of the nanowire.  In the case shown in Fig.~\ref{fig_md_rhombic_3_0.1K}, the twinning-driven reorientation does occur when in fact the initial defect pattern suggests that it should not.  This is not surprising considering that the SSF calculation is done for defect-free, strained structures without any pre-existing stacking faults or twins.  Furthermore, we note that other defect nucleation theories, namely the seminal bulk GSF criteria, would also exhibit this shortcoming if the overall mechanical response is predicted based only on the initial nucleation event.

\begin{figure}\begin{center}
\includegraphics[scale=0.5]{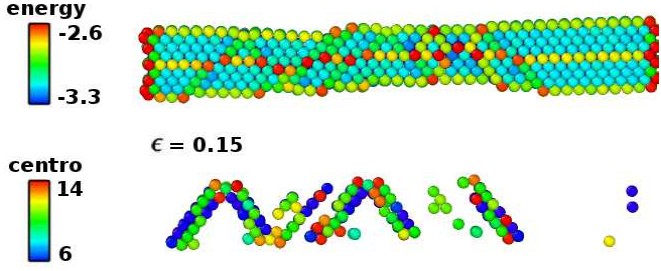}
\caption{(Color online) The evolution of the CuNW with truncation level $n=4$ from MD simulation at 0.1~{K}. Atoms are colored by the potential energy in the top image and by the centrosymmetry parameter in the bottom image. Slip is the only deformation mechanism observed at this temperature.}
\label{fig_md_rhombic_4_0.1K} \end{center} \end{figure}

Fig.~\ref{fig_md_rhombic_4_0.1K} shows the slip-dominated deformation of the rhombic CuNW with truncation level $n=4$ from MD simulation at 0.1~{K}, where the top image is colored according to the potential energy per atom while the bottom image is colored by the centrosymmetry parameter.  The slip deformation and the absence of any twinning is clearly observed in the bottom image.  The MD simulation is consistent with the SSF energy calculation, which predicts the slip deformation to be the preferred pathway in this CuNW.

Before concluding this section, we provide an energetic explanation for the twinning to slip deformation transition in the truncated rhombic CuNWs.  During the deformation of the truncated rhombic CuNWs, the four \{111\} surfaces are reoriented to \{100\} surfaces, while the two \{100\} surfaces are reoriented to \{110\} surfaces.  Furthermore, it is important to note that in both cases, the reoriented surface has a higher surface energy than the initial surface orientation.  Because of this, we calculate the surface energies for the $\{111\}$, $\{100\}$ and $\{110\}$ surfaces to be 1242.3, 1352.6, and 1480.5 mJ/m$^{2}$, which are in agreement with the values published by~\citet{mishinPRB2001}.  Based on these values, it can be seen that there is an energetic penalty of 109.3 mJ/m$^{2}$ to reorient a \{111\} surface into a \{100\} surface.  Similarly, there is a 127.9 mJ/m$^{2}$ energetic penalty for reorienting a \{100\} surface into a \{110\} surface.  In both cases, these energetic differences represent energy barriers that must be overcome by the twinning deformation.  As the truncation level increases, a greater proportion of \{100\} surface area is exposed with a similar reduction in \{111\} surface area.  This is important because, as discussed above, the energetic cost of reorienting a \{111\} surface to \{100\} is lower than the cost of reorienting a \{100\} surface to \{110\}.  This is why the energetic barrier to twinning deformation increases as the truncation of the rhombic CuNWs increases.

In contrast, when slip is the operant deformation mode, no lattice reorientation is observed in the truncated rhombic CuNWs.  In this case, the energy that is added to the system results from the imperfect stacking arrangement in the NW in conjunction with the surface step that results at the free surface due to the stacking fault.  The energetic barrier for this process also changes with increasing truncation, but not as dramatically as it does during the twinning-dominated deformation mode.  As a result, there is a predicted twinning to slip transition in the deformation mechanism as discussed above.

\subsubsection{Discussion on Rhombic NWs}

At this point, it is important to discuss the current SSF predictions for the deformation of rhombic CuNWs with the recent experiments of~\citet{seoNL2011}, who observed the reorientation of $\langle110\rangle/\{111\}$ AuNWs via long-ranged coherent twin propagation.  These experimental results were in contrast to earlier MD predictions for the deformation of $\langle110\rangle/\{111\}$ AuNWs~\cite{parkPRL2005,liangPRB2006}, which predicted that due to the eventual nucleation of full dislocations, the reorientation would not be observed in $\langle110\rangle/\{111\}$ AuNWs.  The relevant implication for the SSF criteria proposed in this work is that it, like the seminal bulk GSF criteria before it, depends crucially on the ability of the interatomic potential to correctly capture the details of the defect nucleation mechanism.  Thus, similar to all other MD simulations, inaccuracies in the interatomic potential will lead to erroneous SSF deformation mechanism predictions.

\begin{figure}\begin{center}
\includegraphics[scale=0.4]{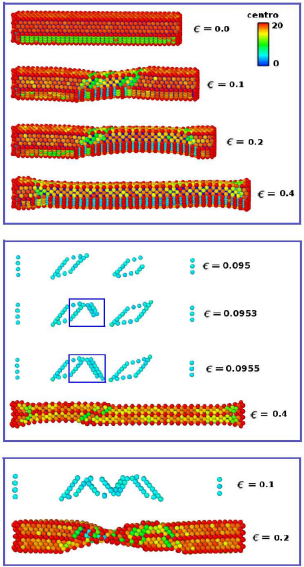}
\caption{(Color online) The stochastic behavior in the evolution of the CuNW with truncation level $n=4$ from MD simulation at 10~{K}. Atoms are colored by the centrosymmetry parameter. The three panels show three independent simulations with different randomly prepared thermal initial conditions. In the top panel, twinning is the only observed deformation mechanism at this temperature. In the middle panel, both twinning and slip deformations are observed, where the growth of a slip fault is highlighted by the blue box. In the bottom panel, only slip deformation occurs. We note much higher ultimate strain in the first two panels than that in the bottom panel, indicating twinning to be the major deformation pathway in the first two panels while slip is the dominant pathway in the bottom panel.}
\label{fig_md_rhombic_4_10K} \end{center} \end{figure}

Another important factor to consider is thermal effects on the SSF predictions.  To study temperature effects on the deformation pathway, we show in Fig.~\ref{fig_md_rhombic_4_10K} the structural evolution from MD simulation for the truncated rhombic CuNW with $n=4$ at 10~{K}. We find that the deformation mechanism indeed exhibits stochastic behavior. The only difference among these three simulations shown in the figure is the initial conditions, which result from different initial distributions of velocities, all satisfying the Boltzmann distribution at 10~{K}. From the simulation shown in the top panel, twinning is found to be the only deformation mechanism, where the twinning leads to the same reorientation previously observed, for example, in Fig.~\ref{fig_md_rhombic_0_0.1K}, and in general for all CuNWs with truncation levels $n\le 2$ at low temperature 0.1~{K}. In the simulation shown in the middle panel, both twinning and slip deformations are found, while in the simulation shown in the bottom panel, only the slip deformation is observed. In the first two cases, the nanowire can undergo a large strain (typically $\epsilon>0.4$), indicating twinning to be the major deformation pathway; while in the last case, the nanowire  fractures at much lower strain (typically $\epsilon <0.2$), which indicates slip as the dominant deformation pathway.

The MD simulation at 10~{K} thus illustrates that the SSF energy calculation cannot predict the deformation pathway at higher temperature, which can be understood from an energetic perspective. 
Fig.~\ref{fig_gsf_rhombic}~(b) shows that the energy barrier between twinning and slip is about $\Delta E=14.2$ mJ/m$^{2}$ at the second unstable stacking fault.  Most of this energy comes from those copper atoms sitting in the two $\{111\}$ planes that are sheared along the $\langle112\rangle$ direction to calculate the SSF energy. The energy barrier can further be distributed onto each copper atom in these two $\{111\}$ planes by multiplying the area per atom in this $\{111\}$ plane: i.e $E_{\rm GSF}=\Delta E \times A_{0} /2$, with $A_{0}=\sqrt{3}a^{2}/4=5.7$~{\AA}$^{2}$, where the factor of 2 is needed to divide the energy barrier onto one of the two shearing $\{111\}$ surfaces. Some simple algebra then gives $E_{\rm GSF}=2.5$~{meV}. 

Besides $E_{\rm GSF}$, there is another important energetic contribution in this system at finite temperature, i.e the thermal vibration energy. According to the equipartition theorem, the vibrational energy of each degree of freedom at temperature $T$ is $k_{B}T/2$, where $k_{B}$ is the Boltzmann constant. The thermal energy for each atom is $E_{T}=3k_{B}T/2$, which is 1.3~{meV} at 10~{K}. We can see that the two energy scales $E_{\rm GSF}$ and $E_{T}$ are on the same order. The thermal vibration reflects a random movement of each copper atom, while the GSF energy curve predicts a deterministic movement. Since the energy scales of these two movements are on the same order, the random movement from thermal vibrations can counteract the deterministic motion imposed in calculating the SSF energy, and is the reason why the SSF prediction fails at higher (i.e. non-zero) temperature.  In contrast, at the low temperature 0.1~{K} that we have used for the majority of our MD simulations, the thermal vibrational energy $E_{T}$ is about two orders smaller than the SSF energy barrier $E_{\rm GSF}$, so the thermal-induced random movement can be safely ignored, resulting in accurate predictions from the SSF energy. 

Finally, we comment on recent experiments that may encourage new interpretations of surface-mediated plasticity.  First, we note the recent experiments of~\citet{sedlmayrAM2012}, who studied AuNWs with the same hexagonal cross section as the CuNWs previously studied by~\citet{richterNL2009}.  In contrast to the brittle fracture observed by~\citet{richterNL2009}, both brittle and ductile behavior was observed by~\citet{sedlmayrAM2012}.  \hsp{In contrast, exclusively brittle fracture was observed in pentagonal AgNWs by both~\citet{zhuPRB2012} and also by~\citet{filleterSMALL2012}.}  These results point to the likelihood of random stochastic effects, such as surface roughness, in yielding seemingly contradictory experimentally observed deformation mechanisms in NWs having the same geometry, or the same geometry and material.  Support for such a stochastic hypothesis was also shown through MD simulations for different initial random velocities (and the same temperature) in Fig.~\ref{fig_md_rhombic_4_10K}.  The key implication is that the SSF results shown in the present work were conducted on nanowires with idealized geometries, no surface roughness, and no stochastic effects due to being at 0~{K}; these differences may lead to different predictions for the deformation mechanisms and mechanical properties than are seen experimentally.

\subsection{\hsp{Universality of the surface stacking fault energy criteria}}


\hsp{In the above, we have demonstrated the agreement between the proposed SSF criteria and the corresponding MD simulations.  However, we considered only partial dislocation slip and twinning as the possible deformation mechanisms and in the calculation of the SSF energy curves.  To demonstrate the universality of the SSF criteria, it is important to demonstrate that it can also predict full dislocation slip, particularly when full dislocation slip is enabled concurrent to both partial dislocation slip and twinning.  The importance of capturing full dislocation slip emerges naturally from our previous study of rhombic nanowires, where earlier MD simulation results by~\citet{liangPRB2006} showed that in contrast to copper, rhombic aluminum nanowires do not undergo the tensile stress-induced reorientation from $\langle110\rangle/\{111\}$ to $\langle100\rangle/\{100\}$ due to the nucleation of full dislocations.}

\hsp{If more than two deformation pathways, i.e. considering full dislocation slip in addition to partial dislocation slip and twinning, is necessary, then SSF criteria can be naturally extended.  Specifically, we calculate the SSF curves for all three deformation mechanisms, and then use the same prediction criteria as before in that the deformation mechanism with the lowest SSF energy at the second unstable stacing fault is the energetically favored pathway for the initial plastic deformation event.}

\begin{figure}\begin{center}
\includegraphics[scale=0.25]{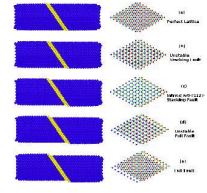}
\caption{(Color online) The formation of a full dislocation fault in the rhombic CuNW with truncation level $n=0$, where the $\{111\}$ plane slips in the $\frac{\vec{a}}{6}[11\bar{2}]+\frac{\vec{a}}{6}[2\bar{1}\bar{1}]$ direction.  The nanowires on the left figures are axially oriented in the $\langle110\rangle$ direction. The viewing direction of the right figures is perpendicular to the shifting $\{111\}$ plane, where A, B, and C layers are displayed by different colors.}
\label{fig_gsf_nw_full} \end{center} \end{figure}

\begin{figure}\begin{center}
\includegraphics[scale=0.7]{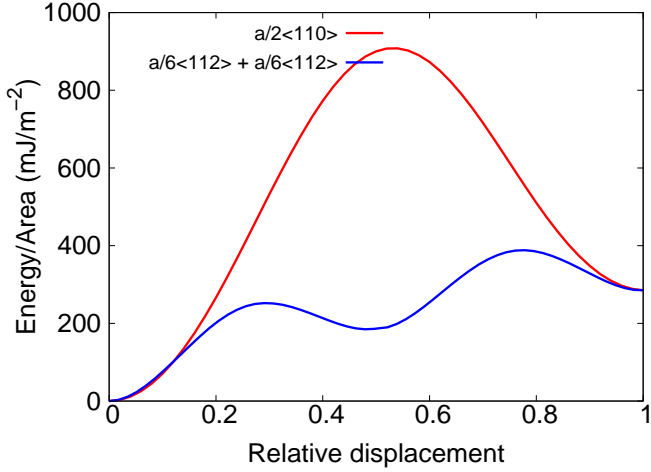}
\caption{(Color online) The SSF energy for full dislocations obtained through the slipping of the $\{111\}$ plane in either the $\frac{\vec{a}}{2}[10\bar{1}]$ or the $\frac{\vec{a}}{6}[11\bar{2}]+\frac{\vec{a}}{6}[2\bar{1}\bar{1}]$ direction.}
\label{fig_gsf_ful_110_112} \end{center} \end{figure}

\hsp{Fig.~\ref{fig_gsf_nw_full} displays the formation of a full dislocation via two partials in the rhombic CuNW with the $\{111\}$ plane slipping in the $\frac{\vec{a}}{6}[11\bar{2}]+\frac{\vec{a}}{6}[2\bar{1}\bar{1}]$ direction.  Of course, the full dislocation can also be formed through the slipping of the $\{111\}$ plane in the  $\frac{\vec{a}}{2}[10\bar{1}]$ direction, as these two deformation paths result in the same full dislocation, since $\frac{\vec{a}}{2}[10\bar{1}]=\frac{\vec{a}}{6}[11\bar{2}]+\frac{\vec{a}}{6}[2\bar{1}\bar{1}]$. Fig.~\ref{fig_gsf_ful_110_112} shows the SSF curves for the full dislocation from these two different paths.  Not surprisingly, it shows that the former deformation pathway comprised of two partials has much lower energy than the single full dislocation, so it will be the favored path.  Therefore, in the following, the SSF energy curves corresponding to the full dislocation are calculated by the summation of two partials, which is obtained by shifting the $\{111\}$ plane in the $\frac{\vec{a}}{6}[11\bar{2}]+\frac{\vec{a}}{6}[2\bar{1}\bar{1}]$ direction.}

\begin{figure}\begin{center}
\includegraphics[scale=1.2]{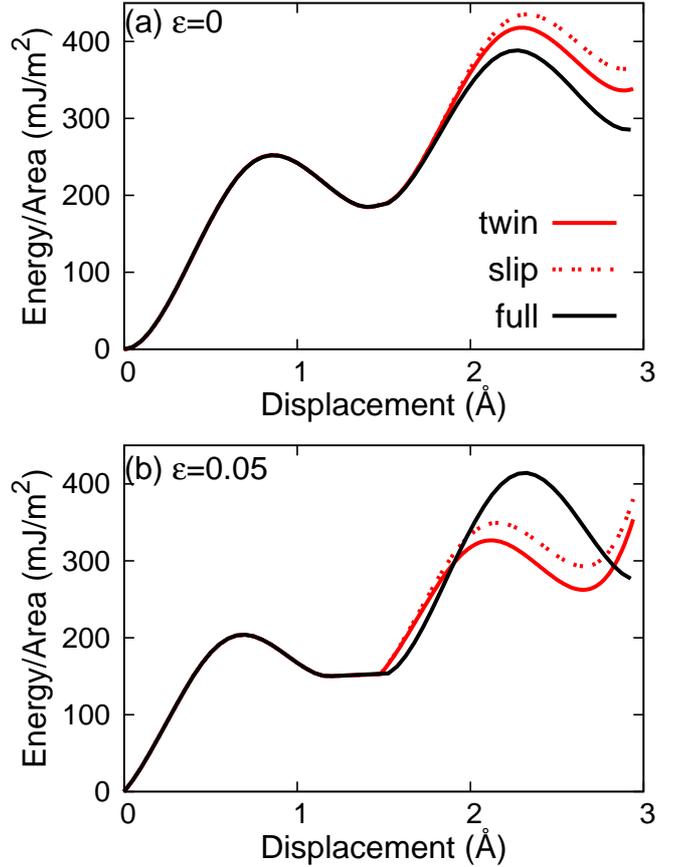}
\caption{(Color online) The SSF energy curves for the rhombic CuNW corresponding to twinning, slip, and full deformations. The SSF energy for structures with strains 0.0 and 0.05 are shown in panels (a) and (b). Note that the full dislocation energy curves becomes less energetically favorable at the higher strain value.}
\label{fig_gsf_slip_twin_full} \end{center} \end{figure}

\hsp{Fig.~\ref{fig_gsf_slip_twin_full} shows the SSF energy curves for the rhombic CuNW corresponding to the twinning deformation, the slip deformation, and the full dislocation.  While all three deformations have the same SSF curve before the first intrinsic stacking fault, they separate afterwards.  Furthermore, as before, we find that tensile strain has an important effect on the SSF energy of the full dislocation. In panel (a), we show that the full dislocation has the lowest SSF energy at the second unstable stacking fault at zero tensile strain.  However, panel (b) shows that at 5\% tensile strain, the SSF energy of the full dislocation becomes the highest among the three deformation mechanisms, which makes it the least energetically favorable.  As has already been discussed, the SSF energy should be applied at a non-zero strain state in order for the most accurate predictions of the favored deformation pathway.  Therefore, the SSF energy curves in Fig.~\ref{fig_gsf_slip_twin_full} demonstrate that twinning is the favored deformation mechanism for rhombic CuNWs with truncation level $n=0$, because the the twinning deformation has the lowest SSF energy at the second unstable intrinsic stacking fault.  This also explains why full dislocations were not observed in our previous MD simulations of rhombic CuNWs.}

\begin{figure}\begin{center}
\includegraphics[scale=1.2]{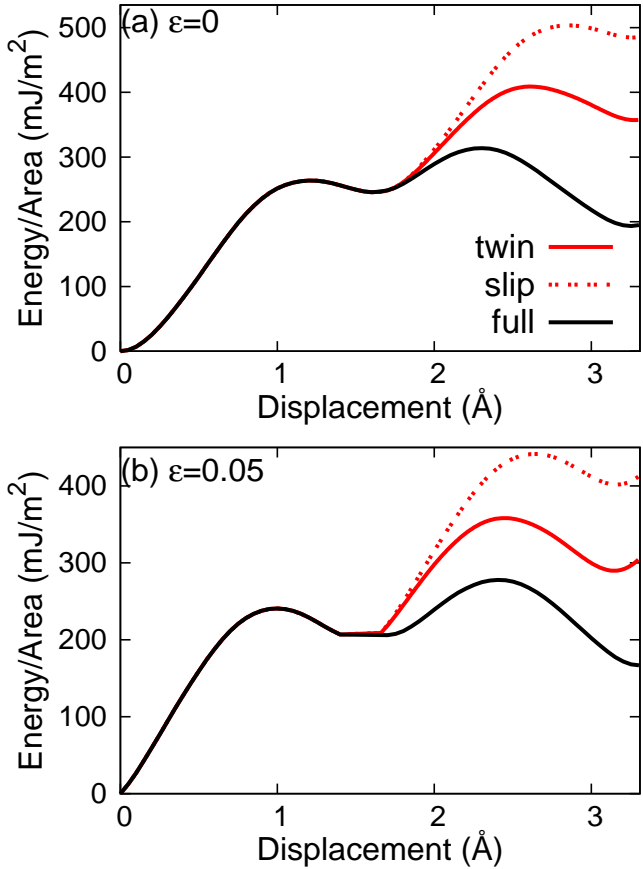}
\caption{(Color online) The SSF energy curves for the rhombic AlNW corresponding to twinning, slip, and full deformations. The SSF energy for structures with strains 0.0 and 0.05 are shown in panels (a) and (b). Note that the SSF curve for full dislocations remains the most favorable as the tensile strain increases.}
\label{fig_al_gsf_slip_twin_full} \end{center} \end{figure}

\begin{figure}\begin{center}
\includegraphics[scale=0.9]{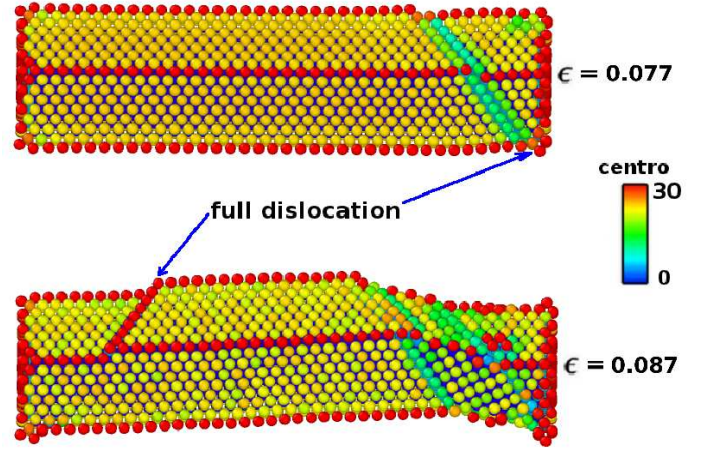}
\caption{(Color online) The evolution of the rhombic AlNW with truncation level $n = 0$ from MD simulation at 0.1 K. Atoms are colored by the centrosymmetry parameter.  The full dislocation is clearly observed at the initial deformation stage.}
\label{fig_al_md_rhombic_0_0.1K} \end{center} \end{figure}

\hsp{To demonstrate that the SSF calculations can distinguish between the three potential deformation mechanisms, we performed SSF calculations for rhombic aluminum (Al) NWs, where the AlNWs had the same geometric structure as the rhombic CuNWs (i.e. same number of atoms), with the only difference being the larger lattice constant of a=4.05~\AA for Al, and the usage of a different EAM potential~\cite{mishinPRB1999} that is appropriate for Al.  Fig.~\ref{fig_al_gsf_slip_twin_full}~(a) shows that the SSF energy for the rhombic AlNW is similar as that of the CuNW without strain, in that the full dislocation has the lowest SSF energy curve.  However, at 5\% tensile strain, we find that the full dislocation still has the lowest SSF energy for the rhombic AlNW, which is different from the trend seen for CuNWs in Fig.~\ref{fig_gsf_slip_twin_full}.  From Fig.~\ref{fig_al_gsf_slip_twin_full}~(b), the SSF criteria predicts the full dislocation to be the favored deformation pathway for this rhombic AlNW, in agreement with the MD simulations of~\citet{liangPRB2006}.  Indeed, Fig.~\ref{fig_al_md_rhombic_0_0.1K} demonstrates that, full dislocations are the initial deformation mechanism in rhombic AlNWs under tensile strain at 0.1~K. This result demonstrates that the SSF criteria can distinguish between full dislocation slip, partial dislocation slip, and twinning to predict the correct deformation mechanism at low temperature.}

\begin{figure}\begin{center}
\includegraphics[scale=1.2]{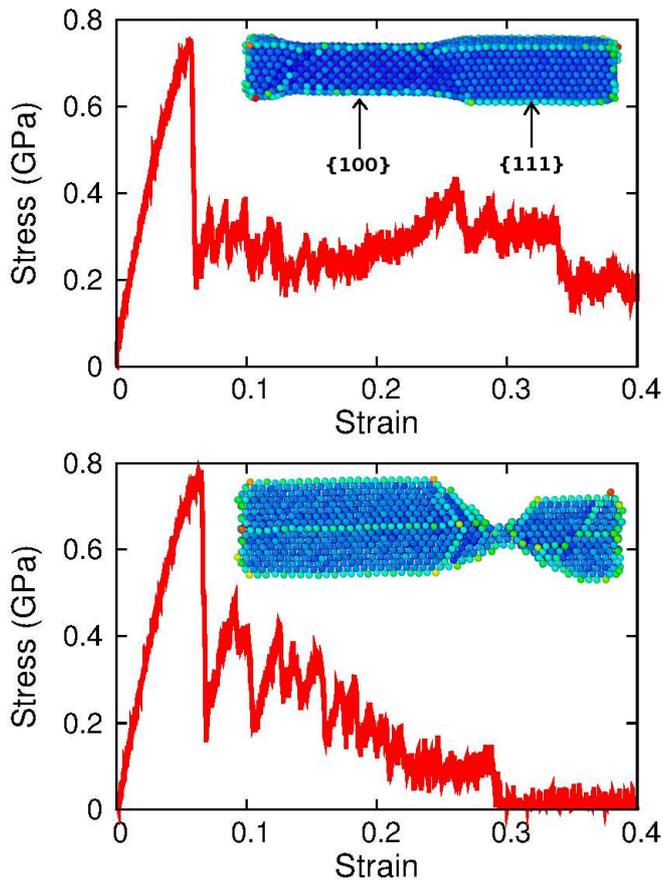}
\caption{(Color online) The strain-stress relation for the rhombic AlNW with truncation level $n=0$ from MD simulation at 300~K. Insets show the configuration at strain 0.2. Atoms are colored by the centrosymmetry parameter. The two panels show two independent simulations with different randomly prepared thermal initial conditions, which shows the stochastic nature of the deformation process.  In the top panel, the NW reorients from the initially rhombic $\langle110\rangle/\{111\}$ NW into a square cross section $\langle100\rangle/\{100\}$ NW through the twinning deformation.  In contrast, necking-induced fracture due to the nucleation of full dislocations is observed in the bottom panel.}
\label{fig_al_md_rhombic_300K} \end{center} \end{figure}

\hsp{Fig.~\ref{fig_al_md_rhombic_300K} shows two MD simulations for the rhombic AlNW at 300~K under tensile strain. The only difference between the two simulations in the figure is their randomly prepared initial velocities, which demonstrates the stochastic nature of the deformation process beyond prediction of the initial deformation mechanism, which the surface SSF criteria captures accurately.  In the top panel, the NW reorients from the initially rhombic $\langle110\rangle/\{111\}$ NW into a square cross section $\langle100\rangle/\{100\}$ NW through the nucleation and propagation of twins. The necking phenomenon induced by the full dislocation is observed in the bottom panel, and is similar to that previously reported by~\citet{liangPRB2006}.  This stochastic effect is similar as what we have observed in Fig.~\ref{fig_md_rhombic_4_10K} for the CuNW. It again illustrates that the SSF criteria is not suitable for higher temperature, where the thermal vibration energy becomes comparable to or larger than the SSF energy barrier.}

\section{Conclusions}

In conclusion, we have proposed a criteria based on the surface stacking fault (SSF) energy to predict the incipient mode of plasticity in surface-dominated nanostructures such as nanowires.  The criteria is based upon whether the unstable twinning energy or unstable slip energy is smaller as determined from the surface stacking fault energy curve.  The validity of the idea was demonstrated by comparison with classical molecular dynamics simulations of nanowires that were previously predicted to exhibit a deformation mechanism transition from slip to twinning based on variations of the nanowire cross sectional geometry.  \hsp{Importantly, the SSF criteria was also shown to accurately distinguish between full dislocation slip, partial dislocation slip and twinning when all three deformation mechanisms are considered concurrently.} The SSF energy criteria was shown to be most accurate at low temperatures, while strain effects were demonstrated to potentially alter the deformation mechanism from slip to twinning for certain nanowire geometries.  

\section{Acknowledgements} JWJ thanks Y. C. Zhang and J. H. Zhao for valuable help in the usage of LAMMPS package. We acknowledge helpful correspondence from J. A. Zimmerman at the Sandia National Laboratory. The work is supported by the Grant Research Foundation (DFG) Germany.  HSP acknowledges support from the NSF grant CMMI-1036460.  \hsp{All authors acknowledge the insightful comments and suggestions of both anonymous reviewers.}


\end{document}